\documentclass[aps,pre,reprint,superscriptaddress,onecolumn,longbibliography,floatfix]{revtex4-2}

\usepackage{empheq}
\usepackage{graphicx}
\usepackage{dcolumn}
\usepackage{bm}

\usepackage[utf8]{inputenc}
\usepackage[T1]{fontenc}
\usepackage{mathptmx}
\usepackage{etoolbox}
\usepackage{xcolor}
\usepackage{mathpazo}
\usepackage{bm}
\usepackage[unicode=true,
 bookmarks=true,bookmarksnumbered=false,bookmarksopen=false,
 breaklinks=false,pdfborder={0 0 0},pdfborderstyle={},backref=false,colorlinks=true]
 {hyperref}
\hypersetup{pdfborderstyle={},pdfborderstyle={},pdfborderstyle={},pdfborderstyle={},pdfborderstyle={},pdfborderstyle={},linkcolor=blue,citecolor=blue}

\usepackage{natbib}
\usepackage{amssymb,amsmath}

\newcommand\beq{\begin{equation}}
\newcommand\eeq{\end{equation}}
\newcommand\beqa{\begin{eqnarray}}
\newcommand\eeqa{\end{eqnarray}}
\newcommand{\nn}{\nonumber\\}
\def\bal#1\eal{\begin{align}#1\end{align}}
    \newcommand{\Tb}{T_{\text{b}}}
    \newcommand{\TA}{T_{\text{A}}}
    \newcommand{\TB}{T_{\text{B}}}
    \newcommand{\Th}{T_{\text{hot}}}
    \newcommand{\Tc}{T_{\text{cold}}}
    \newcommand{\Tw}{T_{\text{warm}}}
    \newcommand{\tw}{t_{\text{w}}}
    \newcommand{\tx}{t_{\times}}
    \newcommand{\tM}{t_{\text{M}}}
       \newcommand{\ww}{\omega}
    \newcommand{\oww}{\widetilde{\ww}}
    
    \newcommand{\otx}{\widetilde{t}_{\times}}
    \newcommand{\otw}{\widetilde{t}_{\text{w}}}
    \newcommand{\otxA}{\widehat{t}_{\times}}
    \newcommand{\Mp}{\text{Mp}}
\newcommand{\taub}{\sigma}
\newcommand{\EE}{\mathcal{E}}
\newcommand{\FF}{\mathcal{F}}
\newcommand{\tauE}{$\tau$-exp}
\newcommand{\ee}{\text{e}}

\makeatletter
\def\@email#1#2{%
 \endgroup
 \patchcmd{\titleblock@produce}
  {\frontmatter@RRAPformat}
  {\frontmatter@RRAPformat{\produce@RRAP{*#1\href{mailto:#2}{#2}}}\frontmatter@RRAPformat}
  {}{}
}%
\makeatother
\begin{document}

\title{The Mpemba effect in the Descartes protocol: a time-delayed Newton's law of cooling approach}
\author{Andr\'es Santos}%
 \affiliation{Departamento de F\'isica, Universidad de Extremadura, E-06006 Badajoz, Spain and Instituto de Computaci\'on Cient\'ifica Avanzada (ICCAEx), Universidad de Extremadura, E-06006 Badajoz, Spain}
\email{andres@unex.es}

\date{\today}

\begin{abstract}
We investigate the direct and inverse Mpemba effects within the framework of the time-delayed Newton's law of cooling by introducing and analyzing the Descartes protocol, a three-reservoir thermal scheme in which each sample undergoes a single-step quench at different times. This protocol enables a transparent separation of the roles of the delay time $\tau$, the waiting time $\tw$, and the normalized warm temperature $\ww$, thus providing a flexible setting to characterize anomalous thermal relaxation. For instantaneous quenches, exact conditions for the existence of the Mpemba effect are obtained as bounds on $\ww$ for given $\tau$ and $\tw$.
Within those bounds, the effect becomes maximal at a specific value $\ww=\oww(\tw)$, and its magnitude is quantified by the extremal value of the temperature-difference function at this optimum.
Accurate and compact approximations for both $\oww(\tw)$ and the maximal magnitude $\Mp(\tw)$ are derived, showing in particular that the absolute maximum at fixed $\tau$ is reached for $\tw=\tau$. A comparison with a previously studied two-reservoir protocol reveals that, despite its additional control parameter, the Descartes protocol yields a smaller maximal magnitude of the effect. The analysis is extended to finite-rate quenches, where strict equality of bath conditions prevents a genuine Mpemba effect, although an approximate one survives when the bath time scale is sufficiently short. The developed framework offers a unified and analytically tractable approach that can be readily applied to other multi-step thermal protocols.
\end{abstract}

\maketitle

\section{Introduction}
\label{sec1}

The Mpemba effect is the counterintuitive phenomenon in which, under certain conditions, a system initially prepared at a higher temperature relaxes to equilibrium faster than an identical system prepared at a lower temperature. Although qualitative observations date back to Aristotle~\cite{aristotle_works_1931}, the effect was brought to modern scientific attention by the experiments of Mpemba and Osborne on water freezing~\cite{MO69}. For decades, it was considered a curiosity specific to water, with explanations invoking evaporation, convection, dissolved gases, or supercooling~\cite{K69,F71,D71,W77,F79,WOB88,A95,M96,ERS08,K09,VM10,VM12,ZHMZZZJS14,VK15,JG15,IC16,SHZMW23}.

In recent years, however, the Mpemba effect has attracted renewed interest as a genuinely nonequilibrium phenomenon of broad relevance~\cite{TBLRV26}. It has been theoretically predicted and experimentally observed in a variety of systems, including granular fluids~\cite{LVPS17,TLLVPS19,BPRR20,MLTVL21,GKG21,GG21,BPR21,BPR22,MS22,MS22b,PSPP23}, spin models~\cite{Betal19,GMLMS21,VD21,TYR23a,TYR23b,D23,GMLMS24}, colloidal suspensions~\cite{KB20,BKC21,CKB21,KCB22,IDLGR24}, Markov jump processes~\cite{LR17,KRHV19,CKB21,BGM21,LLHRW22,TYR23b}, and quantum systems~\cite{CLL21,CTH23,CTH24,MAKP24,Joshietal24,SSMTARO24,WW24,CMA24,YAC24,LZYZ24,SPC25,NE25}. This broader perspective has led to the identification of generalized and inverse Mpemba effects, as well as to a unifying theoretical framework based on the spectral properties of the relaxation dynamics. Consequently, the Mpemba effect is now recognized as a sensitive probe of nonequilibrium relaxation pathways, placing it at the forefront of current research in statistical mechanics and nonequilibrium thermodynamics~\cite{TBLRV26}.

A simple phenomenological framework for studying memory effects in thermal relaxation is provided by the time-delayed Newton's cooling law~\cite{HMMBE23,S24,S25}:
\beq
\label{0.4}
\dot{T}(t)=-\lambda\left[T(t-\tau)-\Tb(t)\right].
\eeq
Here, $\lambda$ is the heat transfer coefficient, $\Tb(t)$ is the time-dependent temperature of the bath, and $0<\tau<\ee^{-1}\simeq 0.368$ is the delay time. In this formulation, the system's temperature evolution depends on its past state rather than solely on its instantaneous deviation from the bath temperature, with the delay parameter $\tau$ playing a central role in capturing nontrivial thermal relaxation dynamics.

The parameter $\tau$ can be interpreted as an effective memory scale accounting for the finite response time of the thermal environment or for the elimination of additional internal degrees of freedom. In this sense, the delayed equation should be viewed as a minimal phenomenological description of non-Markovian relaxation, where the instantaneous rate of change of the temperature depends on its past history. Such a description may arise, for instance, in systems with thermal inertia, in situations where heat exchange is mediated by intermediate subsystems, or in coarse-grained approaches where fast variables have been integrated out.

\begin{figure}
\includegraphics[width=.4\columnwidth]{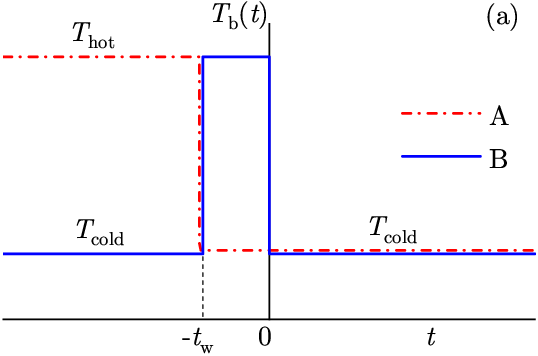}\\
\includegraphics[width=.4\columnwidth]{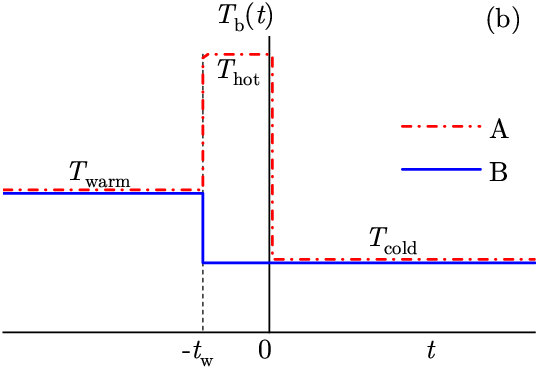}\\
\includegraphics[width=.4\columnwidth]{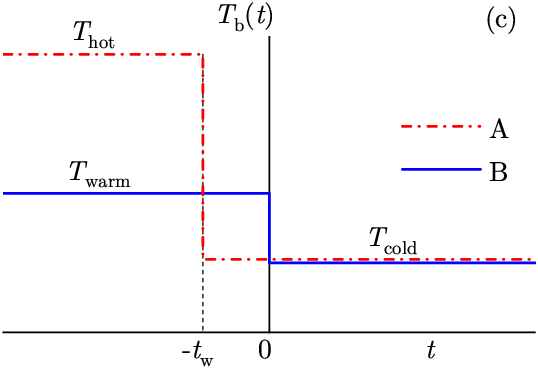}
\caption{Schematic representation of three protocols applied to samples A and B. Panel (a) depicts the two-reservoir protocol of  \cite{S24,S25}, where sample A undergoes a single-step quench at $t=-\tw$ and sample B a double-step quench at $t=-\tw$ and $t=0$. Panel (b) illustrates the Pontus protocol~\cite{NE25,NEDG25}, in which three thermal reservoirs are used: sample B undergoes a single-step quench and sample A a double-step quench. Panel (c) shows the Descartes protocol, where three reservoirs are again used but both samples undergo single-step quenches at $t=-\tw$ (sample A) and $t=0$ (sample B).}
\label{fig0}
\end{figure}

In two recent papers~\cite{S24,S25}, I analyzed the conditions under which equation~\eqref{0.4} predicts a Mpemba effect between two samples (A and B) using only two reservoirs. The protocol was as follows: sample A was first thermalized at the hot temperature $\Th$ and then quenched to the cold temperature $\Tc$ at $t=-\tw$. Sample B was initially thermalized at $\Tc$, quenched to $\Th$ at $t=-\tw$, and finally quenched again to $\Tc$ at $t=0$, resulting in a double-step quench of sample B. Hereafter, this scheme will be referred to as the two-reservoir protocol, to emphasize that only the extreme temperatures $\Th$ and $\Tc$ are employed. Figure~\ref{fig0}(a) schematically illustrates this protocol. \cite{S24} showed that, for a given $\lambda$, the relevant parameter space is two-dimensional (delay time $\tau$ and waiting time $\tw$), and the Mpemba effect occurs only within a specific region of this plane.

A different two-step quench protocol, inspired by Aristotle's account of fishing habits in the Pontus region~\cite{aristotle_works_1931}, has also been proposed~\cite{NE25,NEDG25}. The Pontus protocol, shown in figure~\ref{fig0}(b), begins by equilibrating both samples at a common warm temperature $\Tw$. At time $t=-\tw$, sample A is quenched to $\Th$ and sample B to $\Tc$. Then, at $t=0$, sample A undergoes a second quench to $\Tc$. In contrast to the two-reservoir protocol illustrated in figure~\ref{fig0}(a), the Pontus protocol involves three different reservoirs.

An alternative and simpler protocol is sketched in figure~\ref{fig0}(c). In this case, both samples undergo a single-step quench, and, as in the Pontus protocol, three temperatures ($\Th$, $\Tw$, and $\Tc$) are involved. This preparation of samples A and B is reminiscent of Descartes' prescription~\cite{Descartes1991}:

\begin{quotation}
\noindent
`In order to perform this experiment properly, the water must be allowed to cool down after it has boiled, until it has attained the same degree of coldness as water from a spring. After taking the temperature with a thermometer, you should take some water from a spring and pour equal quantities of the two sorts of water into identical vessels.'
\end{quotation}

In this description, $ \Th$, $ \Tw$, and $ \Tc$ correspond to the boiling temperature, the spring water temperature, and the freezing temperature of water, respectively. In Descartes' original experiment, the waiting time $\tw$ is chosen such that $\TA(0)=\TB(0)=\Tw$. Here, however, we will treat $\tw$ as a control parameter independent of $ \Th$, $ \Tw$, and $ \Tc$.

Throughout this paper, the protocol illustrated in figure~\ref{fig0}(c) will be referred to as the Descartes protocol. The aim is to analyze this protocol within the framework of equation~\eqref{0.4}. As will be seen, the parameter space is three-dimensional: besides the delay time $\tau$ and the waiting time $\tw$, an additional control parameter is the \emph{normalized warm temperature}
\beq
\label{ww}
\ww \equiv \frac{\Tw-\Tc}{\Th-\Tc}.
\eeq
This dimensionless parameter represents the relative position of the intermediate temperature $ \Tw$ between the hot ($ \Th$) and cold ($ \Tc$) temperatures, quantifying how close $\Tw$ is to either extreme. It ranges from $\ww=0$ (if $ \Tw = \Tc$) to $\ww=1$ (if $ \Tw = \Th$).

The paper is organized as follows. The Descartes protocol for the direct Mpemba effect is analyzed in detail in section~\ref{sec2}, while the inverse Mpemba effect is considered in section~\ref{sec3}. Section~\ref{sec4} discusses the influence of finite-rate quenches. The paper concludes with a summary and outlook in section~\ref{sec5}.

\section{The Descartes protocol for the direct Mpemba effect}
\label{sec2}

For simplicity, we henceforth set $\lambda^{-1}=1$ as the unit of time. Since the delay time $\tau$ is inherent to the delayed Newton's cooling law, equation~\eqref{0.4}, rather than to any particular solution, its explicit dependence will be omitted in the notation throughout this paper for clarity.

\subsection{Solution of the time-delayed Newton's cooling equation}

As illustrated in figure~\ref{fig0}(c), in the Descartes cooling protocol, sample A remains in thermal equilibrium with a reservoir at the hot temperature $\Th$ until $t=-\tw$, when it is quenched to the cold temperature $\Tc$. Sample B, on the other hand, is equilibrated with a reservoir at the warm temperature $\Tw$ until $t=0$, when it is also quenched to $\Tc$. If the initial post-quench temperatures satisfy $\TA(0)>\TB(0)$, a Mpemba effect occurs provided that the temperatures cross at some time $\tx>0$, i.e., $\TA(\tx)=\TB(\tx)$, and subsequently satisfy $\TA(t)<\TB(t)$ for $t>\tx$.

From the general solution to equation~\eqref{0.4}~\cite{S25}, the time evolution of the sample temperatures is
\begin{subequations}
\label{TATB_evol}
\beq
\TA(t)=
\begin{cases}
\Th,&t\leq-\tw,\\
\Tc+(\Th-\Tc)\EE(t+\tw),&t\geq -\tw,
\end{cases}
\eeq
\beq
\TB(t)=
\begin{cases}
\Tw,&t\leq 0,\\
\Tc+(\Tw-\Tc)\EE(t),&t\geq 0,
\end{cases}
\eeq
\end{subequations}
where $\EE(t)$ is the quasi-exponential (\tauE) function
\bal
\label{14}
\EE(t)&=1+\sum_{n=0}^{\lfloor t/\tau \rfloor} \frac{(n\tau-t)^{n+1}}{(n+1)!},\quad t\geq 0,\nn
&=
\begin{cases}
1-t,&0\leq t\leq \tau,\\
1-t+\frac{(\tau-t)^2}{2},&\tau\leq t\leq 2\tau,\\
1-t+\frac{(\tau-t)^2}{2}+\frac{(2\tau-t)^3}{3!},&2\tau\leq t\leq 3\tau,\\
\cdots,&\cdots
\end{cases}
\eal
with $\lfloor \cdot \rfloor$ denoting the floor function. Note that
\beq
\label{dotEE}
\partial_t \EE(t) = -\EE(t-\tau),
\eeq
with the convention $\EE(t)=1$ for $t\leq 0$.

For asymptotically long times~\cite{S24,S25},
\beq
\label{AEE0}
\EE(t)\approx \frac{\kappa_0^{-1} \ee^{-\kappa_0 t}}{1-\kappa_0 \tau} - \frac{\kappa_1^{-1} \ee^{-\kappa_1 t}}{\kappa_1 \tau -1},\quad t\gg 1,
\eeq
where $\kappa_0=-\tau^{-1} W_0(-\tau)$ and $\kappa_1=-\tau^{-1} W_{-1}(-\tau)$, with $W_0(z)$ and $W_{-1}(z)$ being the principal and lower branches of the Lambert function~\cite{CGHJK96}. These roots satisfy $1<\kappa_0<e\simeq 2.718<\tau^{-1}<\kappa_1$. For long times, the first term on the right-hand side of equation~\eqref{AEE0} dominates, while the second term improves the accuracy of $\EE(t)$ even at relatively short times.

For $t\geq 0$, equations~\eqref{TATB_evol} yield
\beq
\label{thetaAB}
\theta_{\text{A}}(t)=\EE(t+\tw),\quad \theta_{\text{B}}(t)=\ww\EE(t),\quad t\geq 0,
\eeq
in terms of the normalized temperature
\beq
\label{theta}
\theta(t)\equiv \frac{T(t)-\Tc}{\Th-\Tc}.
\eeq

\begin{figure}
\includegraphics[width=0.4\columnwidth]{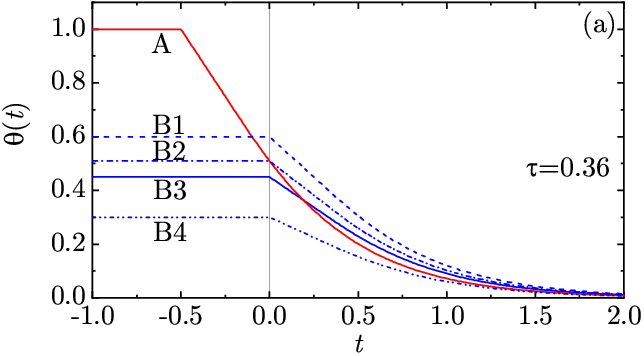}\\
\includegraphics[width=0.4\columnwidth]{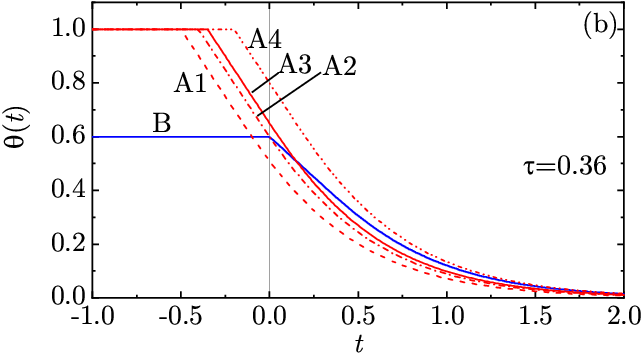}
\caption{Normalized temperature $\theta(t)$ for different samples of type A and type B with a delay time $\tau=0.36$. In panel (a), the waiting time is $\tw=0.5$ for sample A, while the normalized warm temperature is $\ww=0.60$, $0.51$, $0.45$, and $0.30$ in samples B1, B2, B3, and B4, respectively. In panel (b), the normalized warm temperature is $\ww=0.60$ for sample B, while the waiting time is $\tw=0.50$, $0.40$, $0.35$, and $0.20$ in samples A1, A2, A3, and A4, respectively.}
\label{fig2}
\end{figure}

As an example, figure~\ref{fig2} shows the evolution of $\theta(t)$ for different choices of samples of types A and B. The genuine Descartes protocol, in which $\TA(0)=\TB(0)=\Tw$, corresponds to the pairs (A,B2) and (A2,B). The Mpemba crossover is observed in the cases (A,B3) and (A3,B).

In the cases (A,B1) and (A1,B), the waiting time and/or the warm temperature are sufficiently large that $\TA(0)<\TB(0)$, preventing a Mpemba effect. Conversely, in the cases (A,B4) and (A4,B), the waiting time and/or the warm temperature are too small to allow sample A to catch up with sample B.

To analyze the conditions for the Mpemba effect, we  define the difference function
\beq
\label{23}
\Delta(t;\tw,\ww)\equiv \theta_{\text{A}}(t)-\theta_{\text{B}}(t), \quad t\geq 0.
\eeq
From equation~\eqref{thetaAB}, we then have
\beq
\label{Delta}
\Delta(t;\tw,\ww) = \EE(t+\tw) - \ww \EE(t),
\eeq
whose time derivative satisfies
\beq
\label{dotDelta}
\partial_t \Delta(t;\tw,\ww) = -\Delta(t-\tau,\tw,\ww).
\eeq

Notably, with the definitions in equations~\eqref{theta} and \eqref{23},
the difference function obtained from the two-reservoir protocol illustrated
in figure~\ref{fig0}(a) coincides exactly with twice the difference function
of the Descartes protocol for the special case $\ww=1/2$. Consequently,
the results on the Mpemba effect reported in  \cite{S24,S25} are
recovered as a particular limit of the present analysis when $\ww=1/2$.

\begin{figure*}
\includegraphics[width=0.34\columnwidth]{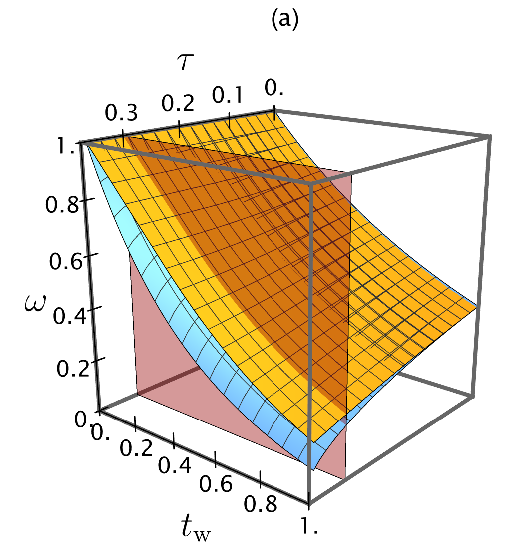}\includegraphics[width=0.34\columnwidth]{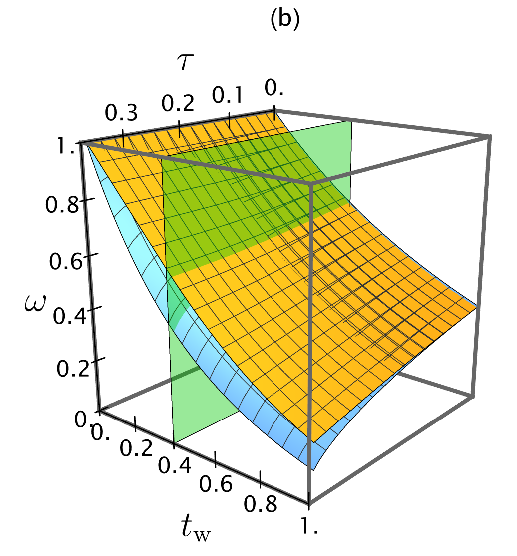}\includegraphics[width=0.34\columnwidth]{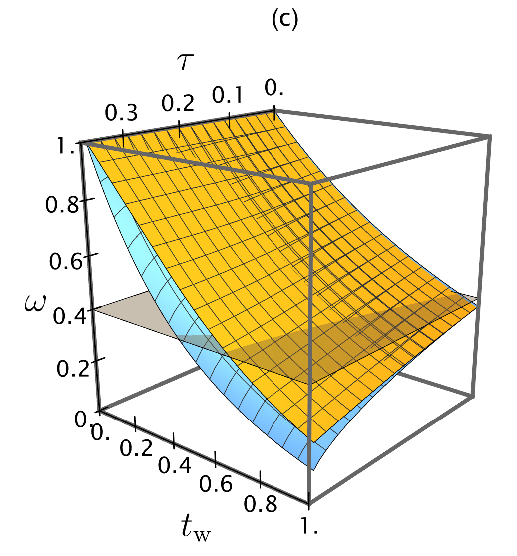}\
\includegraphics[height=0.25\columnwidth]{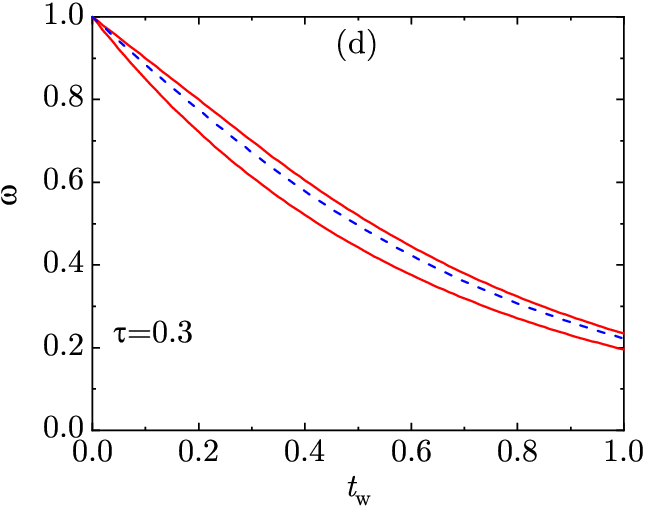}\hspace{0.25cm}\includegraphics[height=0.25\columnwidth]{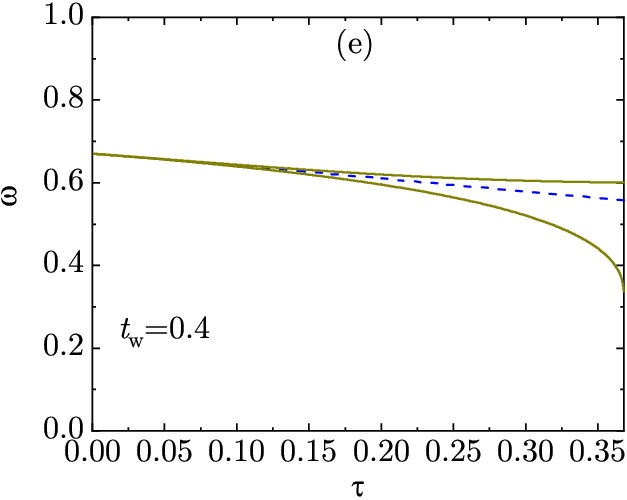}\hspace{0.28cm}\includegraphics[height=0.25\columnwidth]{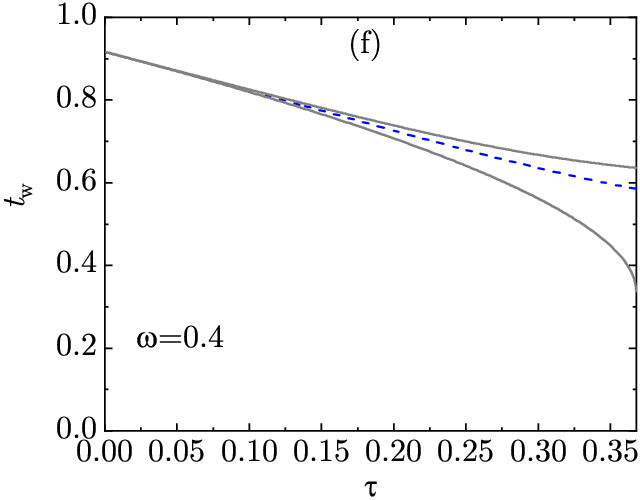}
\caption{Phase space for the direct Mpemba effect. In the top panels (a), (b), (c), the upper surface corresponds to the locus $\ww = \EE(\tw)$, while the lower surface represents $\ww = \ee^{-\kappa_0 \tw}$. The planes shown are (a) $\tau = 0.3$, (b) $\tw = 0.4$, and (c) $\ww = 0.4$. The region $\ww > \EE(\tw)$ is characterized by $\Delta(t \geq 0;\tw, \ww) < 0$, whereas $\Delta(t \geq 0;\tw, \ww) > 0$ holds in the region $\ww < \ee^{-\kappa_0 \tw}$. Consequently, a Mpemba effect occurs if and only if condition \eqref{Mp_cond} is satisfied.
Panels (d), (e), (f) show two-dimensional cross-sections of the three-dimensional phase space at $\tau = 0.3$, $\tw = 0.4$, and $\ww = 0.4$, respectively. Dashed lines indicate the loci where the Mpemba effect is maximal (see section~\ref{sec2C} and equation~\eqref{MaximME}).}
\label{fig1}
\end{figure*}

\subsection{Conditions for the Mpemba effect}

We now analyze the conditions for the existence of a Mpemba effect. The first basic requirement is $\Delta(0;\tw,\ww)>0$, as occurs for the pairs (A,B3), (A,B4), (A3,B), and (A4,B) in figure~\ref{fig2}. The second requirement is that $\Delta(t;\tw,\ww)<0$ after a certain crossover time $\tx$, as observed for the pairs (A,B3) and (A3,B).

At $t=0$,
\beq
\label{Delta00}
\Delta(0;\tw,\ww) = \EE(\tw) - \ww.
\eeq
Hence, $\Delta(0;\tw,\ww) > 0$ if $\ww < \EE(\tw)$.

On the other hand, for long times, using equation~\eqref{AEE0},
\beq
\label{ADelta0}
\Delta(t;\tw,\ww) \approx \left(\ee^{-\kappa_0 \tw}-\ww\right)\frac{\kappa_0^{-1} \ee^{-\kappa_0 t}}{1-\kappa_0 \tau}, \quad t\gg 1.
\eeq
Therefore, $\Delta(t;\tw,\ww)$ approaches zero from below if $\ww > \ee^{-\kappa_0 \tw}$.

In summary, a Mpemba effect occurs if and only if the normalized warm temperature $\ww$ lies within the interval
\beq
\label{Mp_cond}
\ee^{-\kappa_0 \tw} < \ww < \EE(\tw).
\eeq
Using the relation $\EE(t) = 1 - t$ for $t<\tau$, this condition reduces, for small $\tw$, to
\beq
\label{Mp_condLowtw}
\tw < 1-\ww < \kappa_0 \tw, \quad \tw \ll 1.
\eeq

If condition \eqref{Mp_cond} is satisfied, the difference function $\Delta(t;\tw,\ww)$ must vanish at a crossover time $t = \tx(\tw,\ww)$, defined by
\beq
\label{CrossoverTime}
\ww = \frac{\EE(\tx+\tw)}{\EE(\tx)}.
\eeq
This equation must generally be solved numerically. In the limiting case where $\ww$ is slightly below its upper bound $\EE(\tw)$, an approximate solution is
\beq
\label{txHighw}
\tx(\tw,\ww) \approx \frac{\EE(\tw)}{\EE(\tw-\tau)-\EE(\tw)} \left[1 - \frac{\ww}{\EE(\tw)}\right].
\eeq

The complementary case, where $\ww$ is slightly above its lower bound $\ee^{-\kappa_0 \tw}$, requires retaining both the leading and subleading terms in equation~\eqref{AEE0}. After some algebra, the resulting approximation is
\beq
\label{txLoww}
\tx(\tw,\ww) \approx \frac{1}{\kappa_1 - \kappa_0} \ln\left[ \frac{1-\kappa_0 \tau}{\kappa_1 \tau - 1} \frac{1 - \ee^{-(\kappa_1-\kappa_0)\tw}}{\ww \ee^{\kappa_0 \tw} - 1} \right] - \tau.
\eeq

After the crossover time, $-\Delta(t;\tw,\ww)$ reaches a maximum at a time $\tM(\tw,\ww)$, defined by $\partial_t \Delta(t;\tw,\ww) = 0$. Using equation~\eqref{dotDelta}, we have
\beq
\tM(\tw,\ww) = \tx(\tw,\ww) + \tau.
\eeq

The region in the three-dimensional parameter space ${\tau, \tw, \ww}$ that satisfies equation~\eqref{Mp_cond} is illustrated in figures~\ref{fig1}(a)--(c), while figures~\ref{fig1}(d)--(f) show the corresponding two-dimensional projections obtained by intersecting this 3D space with the planes $\tau = 0.3$, $\tw = 0.4$, and $\ww = 0.4$, respectively.

The upper and lower surfaces in figures~\ref{fig1}(a)--(c) meet along the lines $(\tw, \ww) = (0,1)$ and $(\tau, \ww) = (0, \ee^{-\tw})$. These intersections correspond to the points $(\tw, \ww) = (0,1)$, $(\tau, \ww) = (0,0.67)$, and $(\tau, \tw) = (0,0.92)$ in figures~\ref{fig1}(d)--(f), respectively.

For $\tau = 0.36$ and $\tw = 0.5$, the bounds of the Mpemba interval are $\ww = \ee^{-\kappa_0 \tw} \simeq 0.33$ and $\ww \simeq \EE(\tw) \simeq 0.51$. This explains why a Mpemba effect is observed in figure~\ref{fig2}(a) only for the case (A,B3). Similarly, for $\tau = 0.36$ and $\ww = 0.6$, the existence of a Mpemba effect requires $0.23 < \tw < 0.40$, justifying the crossover seen for the case (A3,B) in figure~\ref{fig2}(b).

\begin{figure}
\includegraphics[width=0.4\columnwidth]{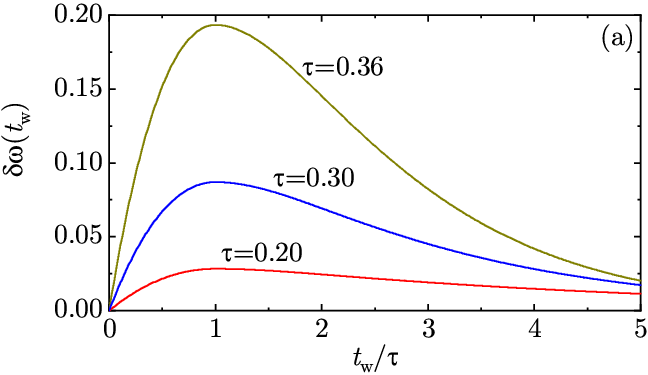}\\
\includegraphics[width=0.4\columnwidth]{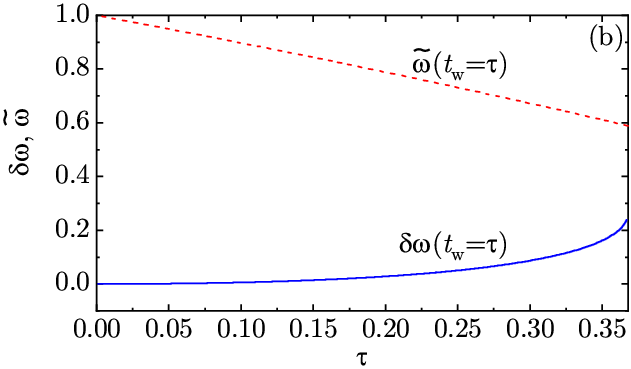}
\caption{(a) Width $\delta\ww(\tw)$ as a function of $\tw/\tau$ for $\tau=0.20$, $0.30$, and $0.36$.
(b) Plot of $\delta\ww(\tw=\tau) = 1 - \tau - \kappa_0^{-1}$ and $\oww(\tw=\tau)$ (see section~\ref{sec2C} and equation~\eqref{owwApprox_tau}) versus $\tau$.}
\label{fig3}
\end{figure}

\subsection{Optimal waiting time}
\label{sec2C}

Figure~\ref{fig1}(d) shows that, for a fixed delay time $\tau$, the width of the $\ww$-interval,
\beq
\delta\ww(\tw) \equiv \EE(\tw) - \ee^{-\kappa_0 \tw},
\eeq
vanishes in the limits $\tw \to 0$ and $\tw \to \infty$. Using equations~\eqref{14} and \eqref{AEE0}, one finds the asymptotic behaviors
\begin{subequations}
\beq
\delta\ww(\tw) \approx (\kappa_0 - 1) \tw, \quad \tw \ll 1,
\eeq
\beq
\label{deltaomega}
\delta\ww(\tw) \approx \left(\frac{\kappa_0^{-1}}{1 - \kappa_0 \tau} - 1\right) \ee^{-\kappa_0 \tw}, \quad \tw \gg 1.
\eeq
\end{subequations}

This implies the existence of an \emph{optimal} waiting time $\tw=\tw^{\rm opt}$ that maximizes the width $\delta\ww(\tw)$. To determine it, note that
\beq
\partial_{\tw} \delta\ww(\tw) = \kappa_0 \ee^{-\kappa_0 \tw} - \EE(\tw - \tau),
\eeq
which vanishes at $\tw = \tau$. Hence, for a given delay time $\tau$, the optimal waiting time is precisely
\beq
\tw^{\rm opt} = \tau.
\eeq

This resonance-like effect is illustrated in figure~\ref{fig3}(a) for $\tau=0.20$, $0.30$, and $0.36$. The peak of $\delta\ww(\tw)$ at $\tw = \tau$ becomes more pronounced as the delay time $\tau$ increases. The height of this peak is
\beq
\delta\ww(\tw=\tau) = 1 - \tau - \kappa_0^{-1},
\eeq
and is plotted in figure~\ref{fig3}(b). Its maximum value, $1 - 2 \ee^{-1} \simeq 0.264$, occurs at $\tau = \tau_{\max} = \ee^{-1}$.

\subsection{Magnitude of the Mpemba effect}
\label{sec2D}

\subsubsection{Maximal effect}

After identifying the Mpemba region in parameter space, the next step is to assess the magnitude of the effect. If $\ww$ is only slightly smaller than $\EE(\tw)$, the Mpemba effect is weak: $\TA(0)$ is only marginally larger than $\TB(0)$, the crossover occurs at very short times (see equation~\eqref{txHighw}), and the temperature curves $\TA(t)$ and $\TB(t)$ remain close throughout the evolution. Conversely, if $\ww$ is only slightly larger than $\ee^{-\kappa_0 \tw}$, the effect is also weak because the crossover occurs at long times (see equation~\eqref{txLoww}), when both $\TA(t)$ and $\TB(t)$ have nearly equilibrated with the bath temperature $\Tc$.

Given two of the three free parameters ($\tau$, $\tw$, and $\ww$), the effect is maximized when the maximum positive and minimum negative values of $\Delta(t;\tw,\ww)$ are equal in magnitude (but opposite in sign), i.e.,
\beq
\Delta(0;\tw, \ww) = -\Delta(\tx(\tw, \ww)+\tau;\tw, \ww),
\eeq
as proposed in  \cite{S24}. This condition for the \emph{maximal} Mpemba effect leads to
\beq
\label{MaximME}
\ww = \frac{\EE(\tw) + \EE(\tx(\tw,\ww) + \tau + \tw)}{1 + \EE(\tx(\tw,\ww) + \tau)}.
\eeq

The curves defined by equation~\eqref{MaximME} are plotted as dashed lines in figures~\ref{fig1}(d)--(f). Notably, the surface described by equation~\eqref{MaximME} lies closer to the upper surface in figures~\ref{fig1}(a)--(c) than to the lower surface.

For given $\tau$ and $\tw$, let $\oww(\tw)$ denote the solution to equation~\eqref{MaximME}, and let $\otx(\tw) \equiv \tx(\tw, \oww(\tw))$ denote the corresponding crossover time. It can be observed that $\otx(\tw) < \tau$, so that, using equation~\eqref{14},
\begin{subequations}
\beq
\EE(\otx(\tw)) = 1 - \otx(\tw),
\eeq
\beq
\EE(\otx(\tw) + \tau) = 1 - \tau - \otx(\tw) + \frac{\otx^2(\tw)}{2}.
\eeq
\end{subequations}
As a consequence, equations~\eqref{CrossoverTime} and \eqref{MaximME} yield
\begin{subequations}
\label{MaximME2&3}
\beq
\label{MaximME2}
\frac{\EE(\otx(\tw) + \tw)}{1 - \otx(\tw)} = \frac{\EE(\tw) + \EE(\otx(\tw) + \tau + \tw)}{2 - \tau - \otx(\tw) + \otx^2(\tw)/2},
\eeq
\beq
\label{MaximME3}
\oww(\tw) = \frac{\EE(\otx(\tw) + \tw)}{1 - \otx(\tw)}.
\eeq
\end{subequations}

Equation \eqref{MaximME2} is a closed equation for $\otx(\tw)$ that can be solved numerically. Substituting the solution into equation~\eqref{MaximME3} then provides the optimal value of the normalized warm temperature, $\oww(\tw)$.

\begin{figure}
\includegraphics[width=0.4\columnwidth]{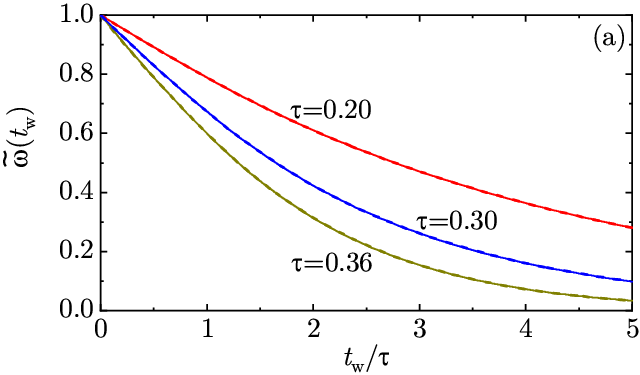}\\
\includegraphics[width=0.4\columnwidth]{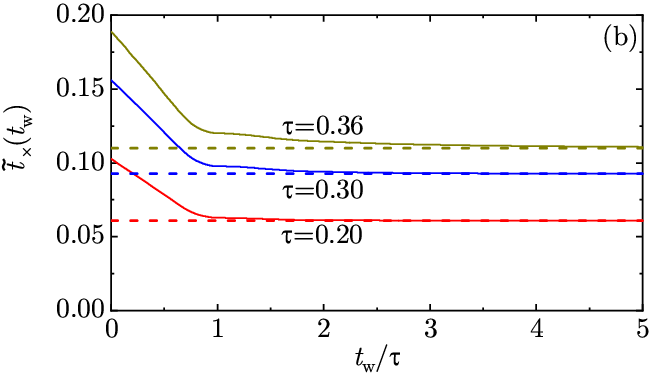}\\
\includegraphics[width=0.4\columnwidth]{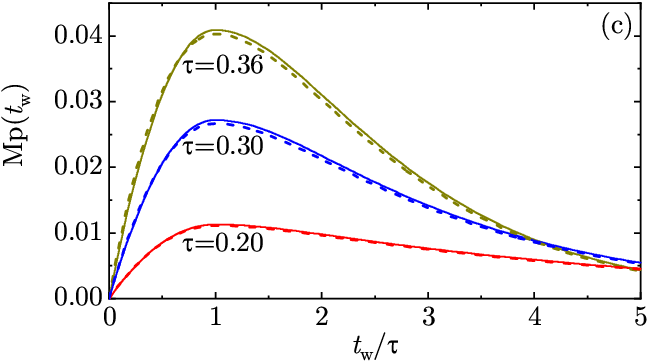}
\caption{Plots of (a) the optimal normalized warm temperature $\oww(\tw)$, (b) the corresponding crossover time $\otx(\tw)$, and (c) the magnitude of the Mpemba effect, $\Mp(\tw)$, all as functions of $\tw/\tau$ for $\tau = 0.20$, $0.30$, and $0.36$. In each panel, the dashed lines show the approximate results obtained from equations~\eqref{owwApprox}, \eqref{otxA}, and \eqref{MpApprox}, respectively.}
\label{fig5}
\end{figure}

Figures~\ref{fig5}(a) and \ref{fig5}(b) illustrate the dependence of $\oww(\tw)$ and $\otx(\tw)$, respectively, on the waiting time $\tw$ for $\tau = 0.20$, $0.30$, and $0.36$. As $\tw$ increases, $\oww(\tw)$ decreases toward zero, while $\otx(\tw)$ approaches a well-defined plateau
\beq
\otxA \equiv \lim_{\tw \to \infty} \otx(\tw).
\eeq
Using equations~\eqref{AEE0} and \eqref{deltaomega}, one finds
\begin{subequations}
\beq
\label{otxA}
\frac{\ee^{-\kappa_0 \otxA}}{1 - \otxA} = \frac{1 + \kappa_0^{-1} \ee^{-\kappa_0 \otxA}}{2 - \tau - \otxA + \otxA^2/2},
\eeq
\beq
\label{owwA}
\oww(\tw) \approx \frac{\kappa_0^{-1}}{1 - \kappa_0 \tau} \frac{\ee^{-\kappa_0(\otxA + \tw)}}{1 - \otxA}, \quad \tw \gg 1,
\eeq
\beq
\label{owwA&MpA}
\lim_{\tw \to \infty} \frac{\oww(\tw)}{\delta\ww(\tw)} = \frac{1}{1 - \kappa_0 (1 - \kappa_0 \tau)} \frac{\ee^{-\kappa_0 \otxA}}{1 - \otxA}.
\eeq
\end{subequations}

\subsubsection{Quantifying the magnitude of the effect}

For given $\tau$ and $\tw$, equations~\eqref{MaximME2&3} determine the value $\oww(\tw)$ that maximizes the Mpemba effect in the sense that the minimum negative value of $\Delta(t;\tw, \ww)$ is equal in magnitude to its maximum positive value. We take this common absolute value as a measure of the Mpemba effect, denoted hereafter by $\Mp(\tw)$. Using equation~\eqref{Delta00}, this is simply
\beq
\label{Mp}
\Mp(\tw) = \EE(\tw) - \oww(\tw).
\eeq

Figure~\ref{fig5}(c) shows $\Mp(\tw)$ for $\tau = 0.20$, $0.30$, and $0.36$. The magnitude exhibits a nonmonotonic dependence on the waiting time $\tw$. In the limit $\tw \to \infty$, one finds
\beq
\label{Mp_oww}
\lim_{\tw \to \infty} \frac{\Mp(\tw)}{\oww(\tw)} = (1 - \otxA) \ee^{\kappa_0 \otxA} - 1.
\eeq

A striking qualitative similarity with figure~\ref{fig3}(a) is apparent: both $\delta\ww(\tw)$ and $\Mp(\tw)$ reach their maximum at $\tw = \tau$. To make this connection more precise, note that from equations~\eqref{owwA&MpA} and \eqref{Mp_oww}, one obtains
\beq
\label{Ratio}
\lim_{\tw \to \infty} \frac{\Mp(\tw)}{\delta\ww(\tw)} = \frac{1 - \ee^{-\kappa_0 \otxA}/(1 - \otxA)}{1 - \kappa_0 (1 - \kappa_0 \tau)}.
\eeq
The right-hand side of equation~\eqref{Ratio} provides an excellent approximation for all $\tw$, particularly for smaller values of $\tau$.

A further simplification is obtained by formally taking $\otx \to 0$ on the right-hand side of equation~\eqref{otxA}, which allows the approximation
\beq
\frac{\ee^{-\kappa_0 \otxA}}{1 - \otxA} \to \frac{1 + \kappa_0^{-1}}{2 - \tau}.
\eeq
This leads to the compact expression
\begin{align}
\label{MpApprox}
\Mp(\tw) &\simeq \frac{1 - \tau - \kappa_0^{-1}}{2 - \tau} \frac{\delta \ww(\tw)}{1 - \kappa_0 (1 - \kappa_0 \tau)} \nonumber\\
&= \frac{1 - \tau - \kappa_0^{-1}}{2 - \tau} \frac{\EE(\tw) - \ee^{-\kappa_0 \tw}}{1 - \kappa_0 (1 - \kappa_0 \tau)},
\end{align}
which does not require solving equation~\eqref{otxA}.

Inserting this result into equation~\eqref{Mp} immediately yields an approximate expression for the optimal normalized warm temperature,
\beq
\label{owwApprox}
\oww(\tw) \simeq \EE(\tw) - \frac{1 - \tau - \kappa_0^{-1}}{2 - \tau} \frac{\EE(\tw) - \ee^{-\kappa_0 \tw}}{1 - \kappa_0 (1 - \kappa_0 \tau)}.
\eeq

As seen in figure~\ref{fig5}(a), equation~\eqref{owwApprox} provides an excellent approximation for $\oww(\tw)$, while figure~\ref{fig5}(c) shows that equation~\eqref{MpApprox} accurately captures the magnitude of the Mpemba effect.

Out of all possible waiting times at fixed $\tau$, the maximum of $\Mp(\tw)$ is reached at $\tw = \tau$. In this case, equations~\eqref{MpApprox} and \eqref{owwApprox} simplify to
\begin{subequations}
\label{MpApprox&owwApprox_tau}
\beq
\label{MpApprox_tau}
\Mp(\tw = \tau) \simeq \frac{(1 - \tau - \kappa_0^{-1})^2}{(2 - \tau)[1 - \kappa_0 (1 - \kappa_0 \tau)]},
\eeq
\beq
\label{owwApprox_tau}
\oww(\tw = \tau) \simeq 1 - \tau - \frac{(1 - \tau - \kappa_0^{-1})^2}{(2 - \tau)[1 - \kappa_0 (1 - \kappa_0 \tau)]}.
\eeq
\end{subequations}
Here we have used the properties $\EE(\tau) = 1 - \tau$ and $\ee^{-\kappa_0 \tau} = \kappa_0^{-1}$. The function $\oww(\tw = \tau)$ is plotted in figure~\ref{fig3}(b).

\begin{figure}
\includegraphics[width=0.4\columnwidth]{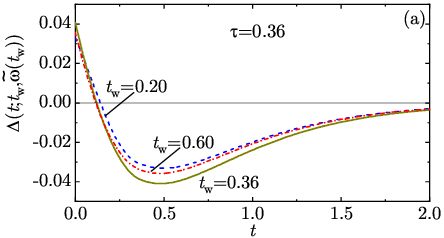}\\
\includegraphics[width=0.4\columnwidth]{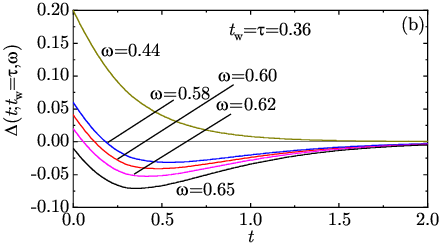}
\caption{(a) $\Delta(t; \tw, \ww)$ for $\tau = 0.36$ and three waiting times: $\tw = 0.20$, $0.36$, and $0.60$. In each case, the normalized warm temperature is set to $\ww = \oww(\tw) \simeq 0.77$, $0.60$, and $0.39$, respectively, ensuring that the maximum positive and minimum negative values of $\Delta(t; \tw, \ww)$ are equal in magnitude. The corresponding magnitudes of the Mpemba effect are approximately $\Mp(\tw) = 0.033$, $0.041$, and $0.036$.
(b) $\Delta(t; \tw, \ww)$ for $\tau = \tw = 0.36$ and five values of the normalized warm temperature: $\ww = 0.44$, $0.58$, $0.60$, $0.62$, and $0.65$. Note that $\oww(\tau = 0.36) \simeq 0.60$.}
\label{fig6}
\end{figure}

The advantage of selecting $\tw = \tau$ to maximize the Mpemba effect is illustrated in figure~\ref{fig6}. Figure~\ref{fig6}(a) shows $\Delta(t; \tw, \ww)$ for a delay time $\tau = 0.36$ and three waiting times ($\tw = 0.20$, $0.36$, and $0.60$), with the normalized warm temperature set to $\ww = \oww(\tw)$ in each case. In all three cases, the maximum positive and minimum negative values of $\Delta(t; \tw, \ww)$ are equal in magnitude, but this common value reaches its highest point when $\tw = \tau$.

Figure~\ref{fig6}(b) presents $\Delta(t; \tw, \ww)$ for $\tau = \tw = 0.36$ and five values of $\ww$: $0.44$, $0.58$, $0.60$, $0.62$, and $0.65$. For these parameters, a Mpemba effect occurs if $\kappa_0^{-1} \simeq 0.45 < \ww < 1 - \tau = 0.64$. This explains why no effect is observed for $\ww = 0.44$ (where $\Delta(t) > 0$ for all $t$) or $\ww = 0.65$ (where $\Delta(t) < 0$ for all $t$). Additionally, $\ww = 0.58$ corresponds to a case where the initial positive $\Delta$ exceeds the absolute value of its minimum, whereas the opposite occurs for $\ww = 0.62$.

Let us summarize the results thus far. For given values of $\tau$ and $\tw$, the existence of a Mpemba effect requires that the normalized warm temperature $\ww$ lies within the interval defined by equation~\eqref{Mp_cond}. Within this interval, the effect is maximal at $\ww=\oww(\tw)$, and the magnitude of this maximal effect is given by $\Mp(\tw)$. Accurate and practical approximations for both quantities are provided by equations~\eqref{MpApprox} and \eqref{owwApprox}.

\begin{figure}
\includegraphics[width=0.4\columnwidth]{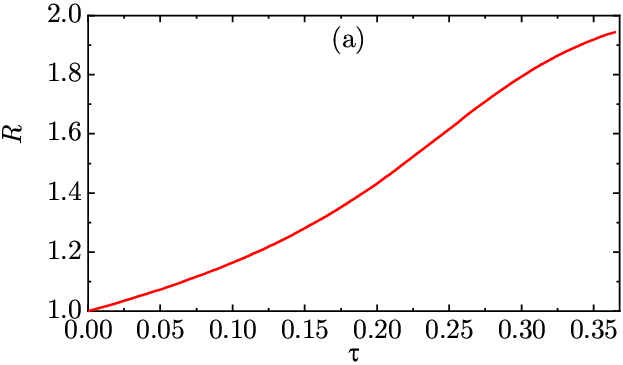}\\
\includegraphics[width=0.4\columnwidth]{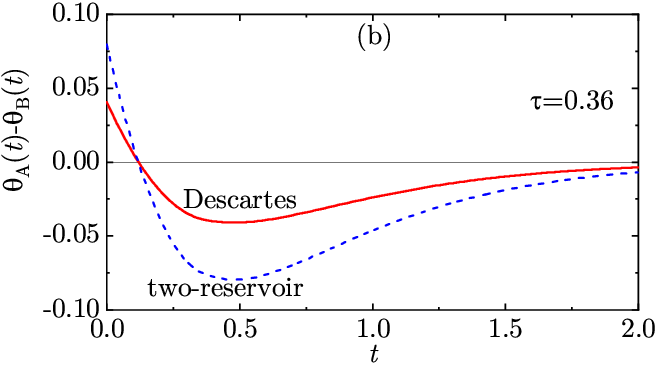}
\caption{(a) Ratio between the magnitude of the Mpemba effect in the two-reservoir
protocol and the maximum magnitude in the Descartes protocol
(see equation~\eqref{R}).
(b) Solid line (Descartes protocol): $\theta_{\text{A}}-\theta_{\text{B}}=\Delta(t;\tw,\ww)$ for $\tau=\tw=0.36$ and $\ww=\oww(\tw)\simeq 0.60$; dashed line (two-reservoir protocol): $\theta_{\text{A}}-\theta_{\text{B}}=2\Delta(t;\tw,\ww)$ for $\tau=0.36$, $\ww=1/2$, and $\tw=\otw\simeq 0.47$. In both cases the temperature difference is normalized with
respect to $\Th-\Tc$.}
\label{fig6.5}
\end{figure}

\subsubsection{Maximum magnitude of the Mpemba effect in the Descartes and two-reservoir protocols}

Given the delay time $\tau$, we have seen that the maximum magnitude of the Mpemba effect in the Descartes protocol is obtained for $\tw=\tau$ and $\ww=\oww(\tw=\tau)$ (see equation~\eqref{MpApprox&owwApprox_tau}).

As noted above, the difference function $\theta_{\text{A}}-\theta_{\text{B}}$ associated with the two-reservoir protocol of figure~\ref{fig0}(a) is exactly twice that of the Descartes protocol with $\ww=1/2$ and the same values of $\tau$ and $\tw$. This comparison is made using the same definition of the
difference function, normalized with respect to the extreme
temperature difference $\Th-\Tc$ (see equation~\eqref{theta}(.
For that value of the normalized warm temperature, the optimal Mpemba effect is obtained with a waiting time $\tw=\otw$, where $\otw$ is the solution to $\oww(\otw)=1/2$.

To compare the magnitudes of the Mpemba effect in the two-reservoir and the Descartes protocols, we introduce the ratio
\beq
\label{R} R=2\frac{\Mp(\tw=\otw)}{\Mp(\tw=\tau)}.
\eeq
It is plotted in figure~\ref{fig6.5}(a) as a function of the delay time $\tau$. We observe that, even though the Descartes protocol includes $\ww$ as an additional control parameter, its maximum Mpemba magnitude is always smaller than that of the two-reservoir protocol with the same delay time. This is further illustrated in figure~\ref{fig6.5}(b) for $\tau=0.36$.

\begin{figure}
  \includegraphics[width=0.4\columnwidth]{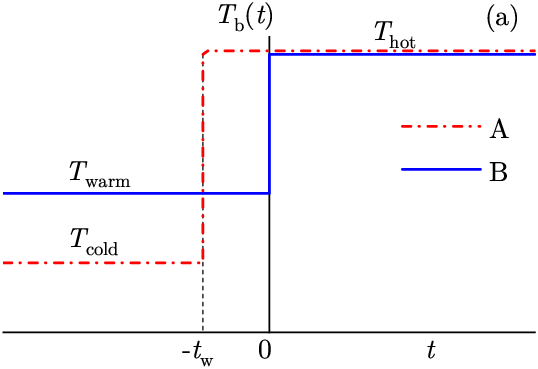}\\
  \includegraphics[width=0.4\columnwidth]{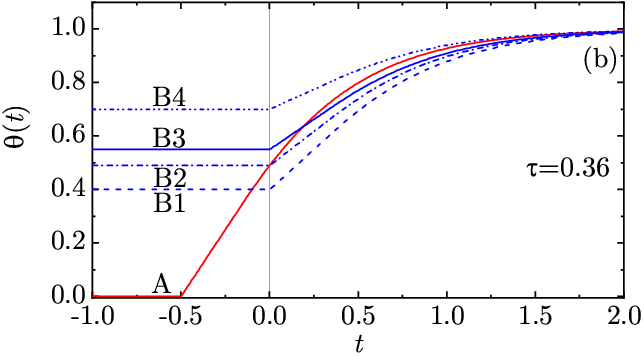}
  \caption{(a) Schematic representation of the Descartes heating protocol.
  (b) Normalized temperature $\theta(t)$ for a sample of type A and different samples of type B with delay time $\tau=0.36$. The waiting time is $\tw=0.5$ for sample A, while the normalized warm temperature is $\ww=0.40$, $0.49$, $0.55$, and $0.70$ for samples B1, B2, B3, and B4, respectively.
  \label{fig7}}
\end{figure}

\section{The Descartes heating protocol for the inverse Mpemba effect}
\label{sec3}
While the Descartes protocol was originally proposed for the conventional cooling scenario, it can be straightforwardly adapted to the heating case. In this variant, sample A remains in thermal equilibrium with a reservoir at a cold temperature $\Tc$ until time $t=-\tw$, when it is quenched to a hot temperature $\Th$. Sample B is equilibrated with a reservoir at an intermediate warm temperature $\Tw$ until $t=0$, when it is also quenched to the same hot temperature $\Th$ as sample A. This heating protocol is sketched in figure~\ref{fig7}(a).
If the initial post-quench temperatures satisfy $\TA(0)<\TB(0)$, an inverse Mpemba effect takes place provided that $\TA(\tx)=\TB(\tx)$ at some crossover time $\tx$, and subsequently $\TA(t)>\TB(t)$ for $t>\tx$.

The equations for the heating process can be obtained from those of the cooling process by performing the formal change $\Tc\leftrightarrow\Th$. In particular, from equation~\eqref{23} one obtains
\beq
\label{23_b}
\frac{\TB(t)-\TA(t)}{\Th-\Tc}=\overline{\Delta}(t;\tw,\ww),\quad t\geq 0,
\eeq
where $\ww$ is still defined by equation~\eqref{ww} and the difference function is now
\bal
\label{Delta_b}
\overline{\Delta}(t;\tw,\ww)=&\Delta(t;\tw,1-\ww)\nn
=&\EE(t+\tw)-(1-\ww)\EE(t).
\eal
The condition for the existence of an inverse Mpemba effect follows from equation~\eqref{Mp_cond} with the replacement $\ww\to1-\ww$:
\beq
\label{IMp_cond}
1-\EE(\tw)<\ww<1-\ee^{-\kappa_0 \tw}.
\eeq

The rest of the results presented in section~\ref{sec2} remain valid, except for the replacement $\ww\to1-\ww$. In particular, figure~\ref{fig7}(b) represents the counterpart of figure~\ref{fig2}(a).
For $\tau=0.36$ and $\tw=0.5$, the bounds of the inverse Mpemba interval are $1-\EE(\tw)\simeq0.49$ and $1-\ee^{-\kappa_0 \tw}\simeq0.67$. This agrees with figure~\ref{fig7}(b), where an inverse Mpemba effect is observed for $\ww=0.55$, but not for $\ww=0.40$ or $\ww=0.70$. The case $\ww=0.49$ corresponds to the heating version of the genuine Descartes protocol, where $\TA(0)=\TB(0)=\Tw$.

\section{Influence of finite-rate quenches}
\label{sec4}

Thus far, we have assumed that, in the Descartes cooling protocol, the bath temperatures experience \emph{instantaneous} quenches from $\Th$ to $\Tc$ at $t=-\tw$ (in the case of sample A) and from $\Tw$ to $\Tc$ at $t=0$ (in the case of sample B). However, it is more realistic to assume that the quenches take place at a finite rate with a characteristic time $\taub$~\cite{S25}, i.e.,
\begin{subequations}
\label{TbA&TbB}
\beq
T_{\text{b,A}}(t)=\begin{cases}
\Th,&t\leq -\tw,\\
\Tc+\left(\Th-\Tc\right)\ee^{-(t+\tw)/\taub},&t\geq -\tw,
\end{cases}
\eeq
\beq
T_{\text{b,B}}(t)=
\begin{cases}
\Tw,&t\leq 0,\\
\Tc+\left(\Tw-\Tc\right)\ee^{-t/\taub},&t\geq 0.
\end{cases}
\eeq
\end{subequations}
Consequently, for $t\geq 0$,
\beq
\label{TbA&TbB_2}
\frac{T_{\text{b,A}}(t)-T_{\text{b,B}}(t)}{\Th-\Tc}=
\left(\ee^{-\tw/\taub}-\ww\right)\ee^{-t/\taub}, \quad t\geq 0.
\eeq
The corresponding time-dependence of the temperature samples is again given by equation~\eqref{TATB_evol}, except that the function $\EE(t)$ is now replaced by~\cite{S25}
\beq
\EE_\taub(t)=\EE(t)+\FF_\taub(t),
\eeq
where the function $\FF_\taub(t)$ is given by
\begin{subequations}
\beq
\FF_\taub(t)=-\sum_{n=0}^{\lfloor{t/\tau}\rfloor}\taub^{n+1}R_n((t-n\tau)/\taub),
\eeq
\beq
R_n(x)\equiv \ee^{-x}-\sum_{k=0}^{n}\frac{(-x)^k}{k!}.
\eeq
\end{subequations}
A practical approximate form for $\EE_\taub(t)$ is
\beq
\label{AEE}
\EE_\taub(t)\approx
\frac{\kappa_0^{-1} \ee^{-\kappa_0 t}}{(1-\kappa_0\taub)(1-\kappa_0\tau)}+
\frac{\taub \ee^{-t/\taub}}{\taub \ee^{\tau/\taub}-1}-\frac{\kappa_1^{-1} \ee^{-\kappa_1 t}}{(1-\kappa_1\taub)(\kappa_1\tau-1)},\quad t\gg 1.
\eeq
This expression reduces to equation~\eqref{AEE0} in the limit of instantaneous quenches ($\taub=0$).
In the long-time limit, the first term dominates over the second one if $\kappa_0\taub<1$, whereas the opposite occurs if $\kappa_0\taub>1$.

The relative temperature difference for $t\geq 0$ is given by a straightforward extension of equation~\eqref{23}:
\begin{subequations}
\beq
\label{23sigma}
\frac{\TA(t)-\TB(t)}{\Th-\Tc}=\Delta_\taub(t;\tw,\ww),\quad t\geq 0,
\eeq
\beq
\label{Delta_sigma}
\Delta_\taub(t;\tw,\ww)=\EE_\taub(t+\tw)-\ww\EE_\taub(t).
\eeq
\end{subequations}
From equation~\eqref{AEE}, we see that the asymptotic form of $\Delta_\taub(t)$ is
\beq
\label{ASS}
\Delta_\taub(t;\tw,\ww)\approx
\frac{\kappa_0^{-1}(\ee^{-\kappa_0\tw}-\ww) \ee^{-\kappa_0 t}}{(1-\kappa_0\taub)(1-\kappa_0\tau)}+
\frac{\taub (\ee^{-\tw/\taub}-\ww)\ee^{-t/\taub}}{\taub \ee^{\tau/\taub}-1}-\frac{\kappa_1^{-1}(\ee^{-\kappa_1\tw}-\ww) \ee^{-\kappa_1 t}}{(1-\kappa_1\taub)(\kappa_1\tau-1)},\quad t\gg 1.
\eeq

Following a reasoning similar to that of section~\ref{sec2}, one finds that the condition for the existence of a direct Mpemba effect changes from equation~\eqref{Mp_cond} to
\beq
\label{Mp_cond_sigma}
\ee^{-\tw \min(\kappa_0,\taub^{-1})} < \ww < \EE_\taub(\tw).
\eeq

Since $\EE_\taub(t)>\EE(t)$, equation~\eqref{Mp_cond_sigma} suggests that the window of values of the normalized warm temperature $\ww$ for which a Mpemba effect exists widens as $\taub$ departs from $0$. However, this is misleading because, as shown by equation~\eqref{TbA&TbB_2}, the environmental conditions for samples A and B are, in general, different if $\taub\neq 0$.

An unbiased Mpemba effect would require $T_{\text{b,A}}(t)=T_{\text{b,B}}(t)$ for $t>0$, but this is only possible (apart from the case $\taub=0$) if the waiting time $\tw$ and the quench characteristic time $\taub$ are coupled by
\beq
\label{twtaub}
\tw=\taub\ln\ww^{-1}.
\eeq
It can be checked that $\ww^{-1}\EE_\taub(\tw=\taub\ln\ww^{-1})$ is always larger than $1$, so that, if $\tw=\taub\ln\ww^{-1}$, the upper bound in equation~\eqref{Mp_cond_sigma} is satisfied for any $\tau$ and $\ww$.
Let us see that, on the other hand, the lower bound is incompatible with equation~\eqref{twtaub}. The first inequality in equation~\eqref{Mp_cond_sigma} is equivalent to $\tw \min(\kappa_0,\taub^{-1})>\ln\ww^{-1}$. Inserting equation~\eqref{twtaub}, this condition becomes $\min(\kappa_0\taub,1)>1$, which is impossible to satisfy.

To reach the same conclusion from an alternative viewpoint, note that the origin of the first inequality in equation~\eqref{Mp_cond_sigma} lies in the condition
\beq
\label{44}
\lim_{t\to\infty}\Delta_\taub(t;\tw,\ww)<0.
\eeq
In the special case of equation~\eqref{twtaub}, the second term on the right-hand side of equation~\eqref{ASS} vanishes identically, so that
\beq
\label{45}
\lim_{t\to\infty}\frac{\Delta_\taub(t;\taub\ln\ww^{-1},\ww)}{\ee^{-\kappa_0 t}}=
\frac{\kappa_0^{-1}}{1-\kappa_0\tau}
\frac{\ww^{\kappa_0\taub}-\ww}{1-\kappa_0\taub}.
\eeq
Since $\ww<1$, the right-hand side of equation~\eqref{45} is strictly positive, thus contradicting the condition in equation~\eqref{44}.

While, strictly speaking, a Mpemba effect is not possible with finite-rate quenches because of the physical requirement of equal bath conditions for both samples, the differences in the environmental conditions become negligible within a tolerable range if $\taub$ is sufficiently small. This approximate equivalence allows the Mpemba effect observed under instantaneous quenches to persist, albeit imperfectly~\cite{S24}.

\section{Conclusions}
\label{sec5}

In this work, the direct and inverse Mpemba effects have been investigated within the framework of the time-delayed Newton's law of cooling by means of the Descartes protocol, a three-reservoir scheme in which each sample undergoes a single thermal quench at different times. This protocol allows one to disentangle the roles of the delay time $\tau$, the waiting time $\tw$, and the normalized warm temperature $\ww$, providing a transparent and flexible setting to analyze the phenomenon.

For instantaneous quenches, we have derived exact conditions for the existence of both the direct and inverse Mpemba effects, expressed as bounds on $\ww$ for given $\tau$ and $\tw$. Within those bounds, the effect becomes maximal at a specific value $\ww=\oww(\tw)$, determined by the condition that the initial difference equals in magnitude the subsequent minimum of $\Delta(t;\tw,\ww)$. At this optimum, the Mpemba magnitude is defined by this common extremal value. Practical and accurate approximations for both $\oww(\tw)$ and the maximal magnitude $\Mp(\tw)$ have been obtained, leading to compact expressions that do not require the numerical solution of transcendental equations. A particularly relevant result is that the absolute maximum of the Mpemba effect at fixed $\tau$ is reached for $\tw=\tau$.

The Descartes protocol has also been compared with the previously studied two-reservoir protocol. Although the Descartes scheme introduces $\ww$ as an additional control parameter, the maximum attainable magnitude of the Mpemba effect is found to be systematically smaller than that of the two-reservoir protocol with the same delay time. This comparison clarifies the relative efficiency of both protocols and highlights the nontrivial interplay between the number of reservoirs and the achievable strength of the effect.

The analysis has been extended to the more realistic situation of finite-rate quenches characterized by a bath time scale $\taub$. While finite-rate quenches formally enlarge the interval of $\ww$ values for which a sign change in $\Delta(t)$ may occur, the requirement of identical bath conditions for both samples after $t=0$ imposes a coupling between $\tw$ and $\taub$ that ultimately precludes a strict Mpemba effect. Nevertheless, when $\taub$ is sufficiently small, the environmental differences become negligible and the behavior approaches that of the instantaneous-quench limit, so that an approximate Mpemba effect can still be observed.

Time-delayed relaxation laws can be effectively realized in systems where the coupling to the thermal environment is not instantaneous, such as feedback-controlled setups, mesoscopic devices with finite environmental response times, or engineered platforms in which delays are deliberately introduced. In these contexts, the delay time $\tau$ is expected to be small but finite compared to the intrinsic relaxation time, which is precisely the regime where the present analysis predicts observable signatures of Mpemba-type behavior. Although the model is intentionally minimal, it provides a controlled framework to assess how memory effects alone can induce anomalous relaxation.

Overall, the Descartes protocol provides a unified and analytically tractable framework to characterize the Mpemba effect and its inverse counterpart, as well as to quantify their magnitude and parametric dependence. An additional conceptual outcome of this work is that Mpemba-type behavior does not require a breakdown of Newton's law of cooling, but can instead emerge naturally within a generalized version that incorporates memory effects through a finite delay time. The methodology developed here can be naturally applied to other multi-step thermal protocols. In particular, a systematic study of the Pontus protocol using the same analytical tools appears as a promising direction for future work, which may further clarify the influence of protocol structure and reservoir sequencing on the emergence and strength of anomalous thermal relaxation.

\begin{acknowledgments}
I acknowledge financial support from Grant No.~PID2024-156352NB-I00 funded by MCIU/AEI/10.13039/501100011033 and by ERDF/EU, and from Grant No.~GR24022 funded by the Junta de Extremadura (Spain).
\end{acknowledgments}

\section*{Data availability statement}
The data that support the findings of this study are openly available at the following URL/DOI:\\ \href{https://github.com/AndresSantosR/The-Mpemba-effect-in-the-Descartes-protocol-A-time-delayed-Newton-s-law-of-cooling-approach}{https://github.com/AndresSantosR/The-Mpemba-effect-in-the-Descartes-protocol-A-time-delayed-Newton-s-law-of-cooling-approach} \cite{S26}.

\bibliography{C:/AA_D/Dropbox/Mis_Dropcumentos/bib_files/Granular}

\end{document}